\input phyzzx.tex
\input epsf

\def\pa{\partial}

\def\caption#1{\vskip 0.1in\centerline{\vbox{\hsize 4.8in\noindent
     \tenpoint\baselineskip=16pt\strut #1\strut}}}
\def\a{\alpha}

\def\g{\gamma}
\def\d{\delta}

\def\et{\eta}

\def\k{\kappa}
\def\l{\lambda}
\def\m{\mu}
\def\n{\nu}
\def\x{\xi}
\def\r{\rho}

\def\s{\sigma}
\def\t{\tau}

\def\ph{\phi}
\def\vp{\varphi}

\def\G{\Gamma}

\def\S{\Sigma}

\def\O{\Omega}
\def\o{\over}
\def\p{\partial}

\def\p{\partial}

\def\g{\gamma}
\def\ap{\alpha_+}

\def\a{\alpha}
\def\G{\Gamma }

\def\e{\varepsilon}
\def\no{\noindent}
\def\inbar{\,\vrule height1.5ex width.4pt depth0pt}
\def\IC{\relax\hbox{$\inbar\kern-.3em{\rm C}$}}
\def\IN{\relax{\rm I\kern-.18em N}}
\def\IR{\relax{\rm I\kern-.18em R}}
\font\cmss=cmss10 \font\cmsss=cmss10 at 7pt
\def\IZ{\relax\ifmmode\mathchoice
{\hbox{\cmss Z\kern-.4em Z}}{\hbox{\cmss Z\kern-.4em Z}}
{\lower.9pt\hbox{\cmsss Z\kern-.4em Z}} {\lower1.2pt\hbox{\cmsss
Z\kern-.4em Z}}\else{\cmss Z\kern-.4em Z}\fi}

\def\tria{\triangleright}
\def\bbox{{\tilde \sqcap}\!\! \!\!\!\!\!\;\sqcup}

\def\na{\nabla}

\def\o{\over}

\def\frac#1#2{{#1 \over #2}}
\def\ot{\otimes}
\def\I{{ I}}
\def\dg{\dagger}
\def\bst{/\!\!\!}
\def\bsf{/\!\!\!\!}
\def\tl{{\tilde \lambda}}
\def\te{{\tilde \varepsilon}}
\def\tg{{\tilde g}}
\def\ap{\alpha_+}
\def\am{\alpha_-}
\def\ba{{\bar a}}
\def\bb{{\bar b}}
\def\ta{\tau_1}
\def\tb{\tau_2}

\def\pot{{\phi\over 2}}

\Ref\BeckerGJ{ K.~Becker and M.~Becker, ``M-Theory on
Eight-Manifolds,'' Nucl.\ Phys.\ B {\bf 477}, 155 (1996),
hep-th/9605053.}

\Ref\BeckerPM{K.~Becker and M.~Becker, ``Supersymmetry breaking,
M-theory and fluxes,'' JHEP {\bf 0107}, 038 (2001),
hep-th/0107044.}

\Ref\BeckerSX{ K.~Becker and K.~Dasgupta, ``Heterotic strings with
torsion,'' JHEP {\bf 0211}, 006 (2002), hep-th/0209077, K.~Becker,
M.~Becker, K.~Dasgupta and P.~S.~Green, ``Compactifications of
heterotic theory on non-K\"a hler complex manifolds. I,'' JHEP
{\bf 0304}, 007 (2003), hep-th/0301161, K.~Becker, M.~Becker,
P.~S.~Green, K.~Dasgupta and E.~Sharpe, ``Compactifications of
heterotic strings on non-Kaehler complex manifolds II,'' Nucl.\
Phys.\ B {\bf 678}, 19 (2004), hep-th/0310058.}

\Ref\BlumenhagenVR{ R.~Blumenhagen, D.~Lust and T.~R.~Taylor,
``Moduli stabilization in chiral type IIB orientifold models with
fluxes,'' Nucl.\ Phys.\ B {\bf 663}, 319 (2003), hep-th/0303016.}

\Ref{\chs}{ C.~G.~Callan, J.~A.~Harvey and A.~Strominger, ``World
sheet approach to heterotic instantons and solitons,'' Nucl.\
Phys.\ B {\bf 359}, 611 (1991);
C.~G.~Callan, J.~A.~Harvey and A.~Strominger, ``Supersymmetric
string solitons,'' arXiv:hep-th/9112030. }

\Ref{\chsb}{ C.~G.~Callan, J.~A.~Harvey and A.~Strominger,
``Worldbrane actions for string solitons,'' Nucl.\ Phys.\ B {\bf
367}, 60 (1991). }

\Ref\CardosoHD{ G.~L.~Cardoso, G.~Curio, G.~Dall'Agata, D.~Lust,
P.~Manousselis and G.~Zoupanos, ``Non-Kaehler string backgrounds
and their five torsion classes,'' Nucl.\ Phys.\ B {\bf 652}, 5
(2003), hep-th/0211118.}

\Ref\chan{C.~S.~Chan, P.~L.~Paul and H.~Verlinde, ``A note on
warped string compactification,'' Nucl.\ Phys.\ B {\bf 581}, 156
(2000), hep-th/0003236.}

\Ref{\DasguptaSS}{ K.~Dasgupta, G.~Rajesh and S.~Sethi, ``M
theory, orientifolds and G-flux,'' JHEP {\bf 9908}, 023 (1999),
hep-th/9908088.}

\Ref\DouglasUM{ M.~R.~Douglas, ``The statistics of string / M
theory vacua,'' JHEP {\bf 0305}, 046 (2003), hep-th/0303194,
S.~Ashok and M.~R.~Douglas, ``Counting flux vacua,'' JHEP {\bf
0401}, 060 (2004), hep-th/0307049, F.~Denef and M.~R.~Douglas,
``Distributions of flux vacua,'' JHEP {\bf 0405}, 072 (2004),
hep-th/0404116, M.~R.~Douglas, ``Statistical analysis of the
supersymmetry breaking scale'', hep-th/0405279, F.~Denef,
M.~R.~Douglas and B.~Florea,``Building a better racetrack,'' JHEP
{\bf 0406}, 034 (2004), hep-th/0404257. }

\Ref{\gmw}{ J.~P.~Gauntlett, D.~Martelli and D.~Waldram,
``Superstrings with intrinsic torsion,'' Phys.\ Rev.\ D {\bf 69},
086002 (2004), hep-th/0302158.}

\Ref{\gaunt}{ J.~P.~Gauntlett and S.~Pakis, ``The geometry of D =
11 Killing spinors,'' JHEP {\bf 0304}, 039 (2003),
hep-th/0212008.}

\Ref\GiddingsYU{ S.~B.~Giddings, S.~Kachru and J.~Polchinski,
``Hierarchies from fluxes in string compactifications,'' Phys.\
Rev.\ D {\bf 66}, 106006 (2002), hep-th/0105097.}

\Ref{\grapol}{ M.~Grana and J.~Polchinski, ``Gauge/gravity duals
with holomorphic dilaton,'' Phys.\ Rev.\ D {\bf 65}, 126005
(2002), hep-th/0106014.}

\Ref\GukovYA{S.~Gukov, C.~Vafa and E.~Witten, ``CFT's from
Calabi-Yau four-folds,'' Nucl.\ Phys.\ B {\bf 584}, 69 (2000)
[Erratum-ibid.\ B {\bf 608}, 477 (2001)], hep-th/9906070.}

\Ref{\glmw}{ S.~Gurrieri, J.~Louis, A.~Micu and D.~Waldram,
``Mirror symmetry in generalized Calabi-Yau compactifications,''
Nucl.\ Phys.\ B {\bf 654}, 61 (2003) arXiv:hep-th/0211102.
}

\Ref{\joyce}{ D.~Joyce, ``Compact manifolds with special
holonomy'', Oxford University Press, Oxford, 2000.}

\Ref\KachruAW{ S.~Kachru, R.~Kallosh, A.~Linde and S.~P.~Trivedi,
``De Sitter vacua in string theory,'' Phys.\ Rev.\ D {\bf 68},
046005 (2003), hep-th/0301240.}

\Ref\KlebanovHB{ I.~R.~Klebanov and M.~J.~Strassler,
``Supergravity and a confining gauge theory: Duality cascades and
$\chi$-SB resolution of naked singularities,'' JHEP {\bf 0008},
052 (2000), hep-th/0007191.}

\Ref\PolchinskiUF{ J.~Polchinski and M.~J.~Strassler, ``The string
dual of a confining four-dimensional gauge
theory,''hep-th/0003136}

\Ref\PolchinskiSM{ J.~Polchinski and A.~Strominger, ``New Vacua
for Type II String Theory,'' Phys.\ Lett.\ B {\bf 388}, 736
(1996), hep-th/9510227.}


\Ref{\schwarz}{ J.~H.~Schwarz, ``Covariant field equations of
chiral N=2 D = 10 supergravity,'' Nucl.\ Phys.\ B {\bf 226}, 269
(1983).}

\Ref{\andy}{ A.~Strominger, ``Superstrings with torsion,'' Nucl.\
Phys.\ B {\bf 274}, 253 (1986).}

\Ref\VanProeyenNI{A.~Van Proeyen, ``Tools for supersymmetry'',
hep-th/9910030.}

\titlepage

\hfill{\vbox{\hbox{hep-th/0410283}}}

\title{\bf A Note on Fluxes in Six-Dimensional String Theory
Backgrounds}

\author{Katrin Becker\foot{katrin@physics.utah.edu} and Li-Sheng Tseng
\foot{tseng@physics.utah.edu}}
\address{\vbox{\hbox{\centerline{Department of Physics}}
\hbox{\centerline{University of Utah}} \hbox{\centerline{Salt Lake
City, UT 84112}}}}

\abstract{We study the structure of warped compactifications of
type IIB string theory to six space-time dimensions.  We find that
the most general four-manifold describing the internal dimensions
is conformal to a K\"ahler manifold, in contrast with the
heterotic case where the four-manifold must be conformally
Calabi-Yau. }

\endpage

\vfill \break

\chapter{Introduction}

Flux compactifications of string theory have attracted much
attention recently because in such backgrounds, many longstanding
questions concerning the connection of string theory to the real
world, are put in a new perspective. To name a few of their
properties, moduli fields can be stabilized at the string tree
level and therefore in a calculable manner [\PolchinskiSM,
\BeckerGJ, \GukovYA, \DasguptaSS, \KachruAW, \BlumenhagenVR].
Supersymmetry can be broken without inducing a large cosmological
constant [\GiddingsYU, \BeckerPM, \KachruAW], and also a large
hierarchy can be induced in a natural way [\GiddingsYU]. In a
different context, flux backgrounds have been shown to provide
gravity dual descriptions to confining gauge theories (see for
example [\KlebanovHB, \PolchinskiUF]).

Because of their importance for phenomenology, most of the recent
work has been concerned with fluxes in four-dimensional
space-times. This paper deals with the less studied case of flux
backgrounds in six dimensions. We will see that in six dimensions,
flux backgrounds are consistent solutions of string theory and
therefore interesting in their own right. Moreover, since the
internal manifolds are four-dimensional, the types of flux
solutions are very much constrained.

For heterotic theory, flux compactifications to six dimensions
were first considered almost twenty years ago [\andy] where a
general analysis of string theory background to
$(10-2n)$-dimensions with tensor fields acquiring an expectation
value was given.  However, it was not until quite recently that
explicit examples of this construction were found and the issue of
moduli stabilization could be addressed  (see for example
[\DasguptaSS, \BeckerSX, \CardosoHD]). In [\andy] it was noticed that
backgrounds with torsion in the connection appear in a natural way
with the torsion being turned on by the three-form tensor field
$H$ satisfying
$$
dH = {\rm tr} R \wedge R -{ 1\o 30} {\rm Tr } F \wedge F~.
\eqn\dhet
$$
In order to preserve supersymmetry, the internal manifold has to
be a complex $n$-manifold and the fundamental form $J_{a \bar b} =
i g_{a \bar b}$ has to satisfy
$$
\pa \bar \pa J = {i \o 30} {\rm Tr} F \wedge F - i {\rm tr} R
\wedge R \qquad {\rm and } \qquad d^\dagger J = i ( \pa - \bar
\pa) \log || \omega ||~. \eqn\hetcond
$$
Here $\omega$ is a holomorphic $(n,0)$ form. The Yang-Mills field
is required to satisfy the Donaldson-Uhlenbeck-Yau equation in a
torsional background. These are the only conditions that have to
be satisfied and many flux compactifications to four dimensions
have been described in the literature.

Specializing to compactifications to six-dimensional Minkowski
space-time, the allowed type of internal four-manifold is
restricted. Indeed, as we will show, supersymmetry can only be
preserved if on the internal manifold, there exists a spinor that
is covariantly constant with respect to a conformally rescaled
metric, i.e. a spinor which satisfies
$$
\nabla'_m \e =0,\eqn\covar
$$
where the covariant derivative is defined with the spin connection
of the rescaled metric. The space-time metric will then be
conformal to a Calabi-Yau two-fold ($K3$ or $T^4$).

In this paper we will study the compactification of type IIB
supergravity to six dimensions in the presence of brane sources.
We study the constraints on the space-time manifold imposed by
non-vanishing fluxes which can be a zero-form and a three-form
tensor field. The two tensor fields can be complex as opposed to
the heterotic case in which they are real. We will see that the
most general four-manifold describing the internal dimensions is
conformal to a K\"ahler manifold, in contrast with the heterotic
case where the four-manifold must be conformally Calabi-Yau.

\chapter{Type IIB string theory compactified to six dimensions}

The ten-dimensional type IIB supersymmetry transformations are
\foot{Our notation mostly follows those of [\schwarz] except we
use a mostly positive signature convention for the metric.  See
the Appendices for a list of our conventions.}
$$
\eqalign{ \d \psi_M &= {1\o \kappa}
\left(\nabla_M-\frac{i}{2}\,Q_M\right) \e + { i \o 480}\,
\Gamma^{N_1 \dots N_5} F_{N_1 \dots N_5} \G_M\, \e\cr &\qquad
-{1\o 96} \left( {\G_M}^{PQR}\, G_{PQR} - 9\, \G^{QR}\,G_{MQR}
\right)
  B^{(10)\star} \e^\star, \cr
  \d \l &= {1\o \kappa}\, \G^M P_M\, B^{(10)\star} \e^\star + {1\o 24}\,
\G^{MNP}
 \, G_{MNP}\, \e.}\eqn\twbv
$$
Here $B^{(10)}$ denotes the ten-dimensional complex conjugation
matrix. This factor does not conventionally appear in the
supersymmetry transformations of the type IIB theory because
usually the Majorana basis is chosen where $B^{(10)}=1$.  But as
will be apparent, this is not an appropriate basis for
dimensionally reducing the fermionic variables to 6+4 dimensions.
The ten-dimensional supersymmetry parameter is complex and
satisfies the Weyl condition
$$
\G^{(10)} \e=\e.\eqn\weylc
$$

We will consider configurations with a six-dimensional Poincar\'e
invariance. The line element is of the form
$$
ds^2=e^{2D} \eta_{\m\n}\, dx^\m dx^\n+ e^{-6D} g_{mn}\,dy^m dy^n~,
\eqn\metric
$$
where Latin indices denote the internal four-dimensional
coordinates while Greek indices denote the six-dimensional
Minkowski space-time coordinates. Moreover, $D=D(y)$ is the warp
factor depending on the coordinates of the internal manifold only.
We have arranged the powers of the warp factor for later
convenience.

In constructing a space-time with six-dimensional Poincar\'e
invariance, we set to zero the following components of the tensor
fields
$$
P_\m = Q_\m = G_{\m MN} = F_{N_1 \dots N_5}=0.\eqn\pqgcond
$$
As a result supersymmetric configurations satisfy
$$
\eqalign{ \left(\nabla_m-\frac{i}{2}Q_m\right) \e\, & - {\kappa \o
24}\, e^{6D}\, {\G_m}^{pqr} G_{pqr} B^{(10)\star} \e^\star  =0~,
\cr
  /  \!\!\! \pa D \,\e & = {\k \o 8}\,e^{6D}\,  / \!\!\!\!
G B^{(10)\star} \e^\star~,\cr  \bsf P B^{(10)\star} \e^\star & =
-{\k\o 4}\,e^{6D}\,\bsf G\,\e~,} \eqn\ai
$$
where we have rescaled the spinor $\e$ according to $\e \to
e^{-3D/2} \e$.
As is implied by the second equation of \ai, a non-constant warp
factor $D(y)$ requires at least one component of $G_{mnp}$ being
non-zero.

Next we decompose the ten-dimensional spinors.  We shall represent
an anti-commuting six-dimensional spinor as a pair of Weyl
spinors, $\x_i$, for $i=1,2$, satisfying the symplectic
Majorana-Weyl condition
$$
\eqalign{
&\G^{(6)}\x_i^\pm = \pm \x_i^\pm, \cr &\e_{ij}B^{(6)\star}
(\x^\pm_j)^\star = \x_i^\pm.\cr} \eqn\aii
$$
Here $\pm$ indicates the six-dimensional chirality.  In four
dimensions we will only impose the Weyl condition.  The commuting
four-dimensional Euclidean spinors are written similarly as
$$
\eqalign{
&\G^{(4)}\eta_i^\pm = \pm \eta_i^\pm, \cr
&\e_{ij}B^{(4)\star}(\eta_j^\pm)^\star=\tilde \eta_i^\pm.\cr}
\eqn\aiii
$$
However, we do not a priori impose the condition $\tilde \eta_i =
\eta_i$.

Under the decomposition of the Lorentz algebra $SO(9,1) \to
SO(5,1) \times SO(4)$, a positive chirality spinor decomposes
according to
$$
{\bf 16}_+ \to ( {\bf 4}_+ , {\bf 2}_+) + ( {\bf 4}_- , {\bf
2}_-)~.
$$
For the supersymmetry parameter $\e$ this decomposition can be
written as
$$
\e = \e^{ij}\left(\x^+_i\otimes  \eta^+_j + \x^-_i
\otimes\eta^-_j\right). \eqn\decomp
$$
First note that the decomposition is invariant under an $SU(2)$
transformation acting on the spinor labels $i$ and $j$.  Secondly,
if we had set $\tilde \eta_i = \eta_i$, then
$B^{(10)\star}\e^\star = \e$ and thus $\e$ becomes a
ten-dimensional Majorana-Weyl spinor.  Indeed, any condition
relating $\et_1$ and $\et_2$ reduces the number of spinor degrees
of freedom by $1/2$.  Now, inserting \decomp\ into \ai\ we see
that we obtain two independent conditions for each of the
six-dimensional chiralities.  So we will relabel the spinors
according to $\x^+_i = \x_i$ and $\eta^+_j=\eta_j$ and work only
with a spinor of positive six-dimensional chirality.  We thus set
$$
\e = \e^{ij} \x_i \otimes \eta_j \qquad {\rm and } \qquad
B^{(10)\star} \e^\star = \e^{ij} \x_i \otimes \tilde
\eta_j~.\eqn\decompt
$$

Inserting the above decomposition into the supersymmetry
variations we obtain that supersymmetry can only be preserved if
and only if the following conditions are satisfied
$$
(\nabla_m-\frac{i}{2}Q_m)\, \eta_j + g_m\, \tilde \eta_j=0~,
\eqn\susya
$$
$$
/\!\!\! \pa D\, \eta_ j = -{1 \o 2} /\!\!\! g\, \tilde \eta_j~,
\eqn\susyb
$$
$$
/\!\!\!\! P\, \tilde \eta_j = /\!\!\! g \,\eta_j~ , \eqn\susyc
$$
Here we have introduced the dualized one-form field $g=-{\k\o
4}\,e^{6D}\, (\star G)\,$, or equivalently,
$$
G_{mnp}=-{4\o \k}\,e^{-6D}\,{\e_{mnp}}^q\,g_q \eqn\gexdef
$$

In the following we will use the conditions \susya-\susyc\ to
determine the form of the supersymmetric background.  First we
note that by using \aiii\ we obtain the following result
$$
\eta^{\dagger\,i} \tilde \eta_j = {1\o 2}\, {\d^i}_j \,v~,
\eqn\etin$$ where $\eta^\dagger = (\eta^\star)^T$ and $v=
{\eta^{\dagger\,k}} \, \tilde \eta_k$ is a complex function. The
covariant derivative on the bilinear spinor is then
$$
\eqalign{ \na_m\left(\et^{\dg\,i} \eta_j\right) &= -{1\o 2}
{\d^i}_j\left(g_m\, v +g^\star_m\, v^\star\right)\cr
&={\d^i}_j\left(\et^{\dagger\,k}\g_m\,\bst\pa D \, \eta_k +
\et^{\dagger\,k}\bst\pa D \,\g_m \, \eta_k\right ) \cr &=2\,
{\d^i}_j\, \pa_m D \,\eta^{\dagger\,k}  \eta_k} \eqn\deta
$$
where in the second line, we have noted the following relation
$$
\eqalign{ g_m\,\et^{\dg\,k}  \tilde \eta_k &= {1\o 2}
\et^{\dg\,k}\left(\g_m\,\bst g + \bst g\,\g_m\right)\tilde
\eta_k\cr &= - \et^{\dg\,k}\,\g_m\,\bst\pa D\,\eta_k -
{\tilde\eta}^{\dg\,k}\,\bst\pa D\,\g_m\,\tilde \eta_k\cr &= -2\,
\et^{\dg\,k}\,\g_m\,\bst\pa D\,\eta_k } \eqn\dexpl
$$
obtained utilizing \susyb\ and its complex conjugated form.
Solving \deta, we find
$$
\et^{\dg\,i} \eta_j = {1\o 2}\, e^{4(D+D_0)}  {\d^i}_j + {A^i}_j
\eqn\etsolve
$$
where $D_0$ is a normalization constant and ${A^i}_j$ is a
constant traceless hermitian matrix.  We can diagonalize on the
$SU(2)$ indices so that
$$
{A^i}_j = {1\o 2}\, e^{4A}\, {(\s_3)^i}_j~.\eqn\adiag
$$
Therefore a nonzero ${A^i}_j$ is equivalent to $\et_1$ and $\et_2$
having different normalizations.  If present, an additional
constraint
$$
D>A-D_0~,\eqn\econst
$$
must be imposed so that $\et^{\dg\,2}\et_2>0$.   This constraint
effectively sets a minimum value for $D$.\foot{For another
scenario where a bound for the warped factor occurs, see [\chan].}

We now renormalize the spinors by defining
$$
\eqalign{ \et_1 &= \a_+\l_1 \cr \et_2 &=\a_-\l_2 }\qquad
\eqalign{\tilde\et_1 &= \a_- \tl_1 \cr \tilde\et_2 &= \a_+\tl_2}
\eqn\adef
$$
where
$$
\a^2_\pm = {1\o 2}\left(e^{4(D+D_0)}\pm e^{4A}\right)~.\eqn\apmdef
$$
The normalized spinors $\l_i$ then satisfy
$$
\l^{\dg\,i}\,\l_j = \d^i_j~. \eqn\lconda
$$
An additional constraint on $\l_i$ comes from the relation
$$
\l^{\dg\,i}\,\tl_j = e^{i\vp} \,\d^i_j~. \eqn\lcondb
$$
where $\vp$ is in general a $y$-dependent function.  This results
from \etin\ and the constraint ${|\l^{\dg\,i}\,\tl_j|}^2=\d^i_j$,
which can be derived by applying the Fierz identity and eq.
(A.12).  Noting that $\l_i$ are two-component spinors, \lconda\
and \lcondb\ together imply that
$$
\tl_i=\e_{ij}B^{(4)\star}(\l_j)^\star=e^{i\vp}\,\l_i~. \eqn\lcondc
$$
The supersymmetry conditions \susya-\susyc\ then become
$$
\eqalign{ (\nabla_m-\frac{i}{2}Q_m+\pa_m\ln\ap) \l_1 &=- {\am \o
\ap} g_m\, e^{i\vp} \l_1 \cr (\nabla_m-\frac{i}{2}Q_m+\pa_m\ln\am)
\l_2 &=- {\ap \o \am} g_m\, e^{i\vp} \l_2} \eqn\sula
$$
$$
\bst \pa D \l_1 = -{1 \o 2} {\am \o \ap} \bst g\,e^{i\vp}
\l_1~,\qquad \bst \pa D \l_2 = -{1 \o 2} {\ap \o \am} \bst g\,
e^{i\vp}\l_2 \eqn\sulb
$$
$$
\bsf P\,e^{i\vp} \l_1 = {\ap \o \am} \bst g \,\l_1~,\qquad\, \bsf
P\,e^{i\vp} \l_2 = {\am \o \ap} \bst g \,\l_2~. \eqn\sulc
$$

Let us clarify the relationships of the various fields above.
First, by considering the expression $(g_m\,
e^{i\vp}\l^{\dg\,1}\l_1 +g^\star_m\,e^{-i\vp} \l^{\dg\,1}\l_1)$
and performing a calculation similar to \dexpl, we obtain the
relation
$$
\eqalign{
g_m\,e^{i\vp}+g^\star_m\,e^{-i\vp}&=-2\,\pa_mD\left({\ap\o\am}+{\am\o\ap}\right)\cr
&= -2\,{\ap\o\am}\,\pa_m\ln\ap = -2\,{\am\o\ap}\,\pa_m\ln\am}.
\eqn\gdrel
$$
The above relation can also be derived simply from the constraint
$\na_m(\l^{\dg\,i}\l_j)=0$. Second, note that \lcondc\ explicitly
relates $\l_1$ with $\l_2$ up to a phase factor $\vp$ that must be
determined.  We can find the variation of $\vp$ by requiring that
the two equations of \sula\ are equivalent given \lcondc.  This
results in
$$
\p_m\vp = -Q_m +{1 \o 2i}\left({\ap\o\am}+{\am\o
\ap}\right)(g_m\,e^{i\vp}-g^\star_m\,e^{-i\vp})~, \eqn\dph
$$
having applied \gdrel\ in the calculation.  Now, with \gdrel\ and
\dph, we can simplify \sula\ further to
$$
\eqalign{ \na_m \l_1&= \left(-{i\o 2}\,\p_m\vp + {i\o 2}\,
W_m\right)\l_1\cr \na_m \l_2&= \left(-{i\o 2}\,\p_m\vp - {i\o 2}\,
W_m\right)\l_2} \eqn\suls
$$
where
$$
W_m={1\o 2i}
\left({\ap\o\am}-{\am\o\ap}\right)(g_m\,e^{i\vp}-g^\star_m\,e^{-i\vp}).
\eqn\wdef
$$

We now proceed to discuss the complex structure of the
four-manifold. Consider the triplet of almost complex structures
$$
{(J_A)_m}^n={i\o 2}\,{(\s_A)^j}_i\, \l^{\dg\,i}\, {\g_m}^n\,
\l_j~. \eqn\hka
$$
for $A=1,2,3$.  Using (A.12), it can be shown that
$$
{(J_A)_m}^n\,{(J_B)_n}^p = -\d_{AB}\,{\d_m}^p +
{\e_{AB}}^C\,{(J_C)_m}^p~. \eqn\hkb
$$
Furthermore, \hka\ and \hkb\ imply the metric $g_{mn}$ on the
four-manifold is hermitian with respect to each of the above
almost complex structures; that is,
$$
{(J_A)_m}^k\,{(J_A)_n}^l\,g_{kl}=g_{mn}\qquad {\rm for~}\qquad
A=1,2,3~. \eqn\hkc
$$
The above three equations taken together define an almost
hyperk\"ahler structure on the four-manifold  (see for example
[\joyce]).  To be specific, we will take $J=J_3$ as the almost
complex structure on the four-manifold.  With respect to $J$, we
have the following $(p,q)$-forms
$$
\eqalign{ \O_{mn}&=(J_2 + i J_1)_{mn}\cr J_{mn}&=(J_3)_{mn}\cr
{\bar\O}_{mn}&=(J_2-i J_1)_{mn}}\qquad\qquad
\eqalign{&(2,0)\cr&(1,1)\cr&(0,2)} \eqn\pqform
$$
The covariant derivatives of the Hermitian form, $J$, and the
complex 2-form, $\O$, can be easily obtained using \suls.  We find
$$
\eqalign{ \na_pJ_{mn}&=0~,\cr \na_p\O_{mn}&=-i\,W_p\,\O_{mn}~,}
\eqn\cvhs
$$
where $\O_{mn}=-\l^{\dg\,1}\g_{mn}\,\l_2\,$.\foot{We point out
that in the intrinsic torsion classification of supersymmetric
compactifications (see for example [\glmw, \CardosoHD, \gaunt]),
\cvhs\ implies that the only nonzero torsional class [\gmw] is
$({\cal W}_5)_m = {J_m}^nW_n$.} From \cvhs, we conclude that the
four-manifold is not only complex, but also K\"ahler.  However,
the manifold is Calabi-Yau if and only if
$\na_pJ_{mn}=\na_p\O_{mn}=0$ and in this case the manifold has a
hyperk\"ahler structure.  This condition is satisfied when
$W_m=0$.

The presence of a complex structure allows us to introduce
holomorphic and anti-holomorphic coordinates, $a,b,\ldots$ and
$\ba,\bb,\ldots$, and take ${J_a}^b=i{\d_a}^b$ and
${J_\ba}^\bb=-i{\d_\ba}^\bb$.  The K\"ahler form is related to the
metric by
$$
J_{a\bb}=ig_{a\bb}~.
$$
which implies
$$
\g_a\l_1=\g^\ba\l_1=0 \qquad{\rm and }  \qquad
\g_\ba\l_2=\g^a\l_2=0~. \eqn\gacon
$$
They can be applied to \sulb-\sulc\ to give
$$
g_a=-2\,{\ap\o\am}\,e^{-i\vp}\,\pa_a D~,\qquad
g_\ba=-2\,{\am\o\ap}\,e^{-i\vp}\,\pa_\ba D~,\eqn\gsol
$$
$$
P_a=-2\,\left({\ap\o\am}\right)^2 e^{-2i\vp}\,\pa_a D~,\qquad
P_\ba=-2\,\left({\am\o\ap}\right)^2 e^{-2i\vp}\,\pa_\ba
D~.\eqn\psol
$$
Using both equations in \gsol, we obtain
$$
g_ae^{i\vp}-g_a^\star
e^{-i\vp}=-2\left({\ap\o\am}-{\am\o\ap}\right)\pa_aD~. \eqn\galp
$$

Therefore, if $\a_+ \neq \a_-$ and $\pa_m D\neq 0$ in a region on
the internal four-manifold then Im$[g_me^{i\vp}]\neq0$ and the
background geometry will be conformal to a K\"ahler but
non-Calabi-Yau manifold. The precise relation between $\a_+$ and
$\a_-$ is determined by the constant $A$ which is not fixed by the
present analysis.

In the following we consider the special class of solutions with
$$
C_0=0~,\qquad{\rm and}\qquad {\rm Im}\,[\,g_me^{i\vp}\,]=0~.
\eqn\scs
$$
where $C_0$ is the R-R 1-form and the second condition implying
$\ap=\am$.  From (B.2), we know that in such backgrounds,
$$
Q_m=0~,\qquad{\rm and}\qquad P_m={1\o 2}\,\pa_m\ph~.\eqn\fimp
$$
With \dph-\wdef, these conditions imply
$$
\pa_m\vp=0~,\qquad \na_m\l_1=\na_m\l_2=0~,\qquad
W_m=0~,\eqn\cycond
$$
or that the four-manifold is conformally Calabi-Yau.  Furthermore,
with $P_m$ real, \psol\ implies solutions only for $\vp=0,{\pi\o
2}, \pi, {3\pi\o 2}$.  We therefore have
$$\eqalign{
g_m&=-2\,e^{-i\vp}\,\pa_m D~,\cr P_m&={1\o
2}\,\pa_m\ph=-2\,e^{-2i\vp}\,\pa_m D~,}\qquad\qquad{\rm
for}~~~\vp=0,{\pi\o 2}, \pi, {3\pi\o 2}~.\eqn\spcsol
$$
The sourceless Bianchi identity for $G_{mnp}$ is
$$
\pa_{[m}G_{npq]}=-P_{[m}G^\star_{npq]}~. \eqn\slbian
$$
With $G_{npq}\sim e^{-6D}{\e_{npq}}^rg_r$ as in \gexdef, \slbian\
implies
$$
\eqalign{ 0&=-\pa_{[m}\left(e^{-8D}{\e_{npq]}}^r\pa_rD\right)\cr
&=-\pa_{[m}\left({{\tilde \e}_{npq]}}^{~~~~r}\pa_rD\right)\cr
&=\bbox D} \eqn\dlap
$$
where in the second line, we have rescaled the metric to ${\tilde
g}_{mn}=e^{-8D}g_{mn}$, which leads to $\te_{mnpq}\,{\tilde
g}^{qr}=e^{-8D}\e_{mnpq}g^{qr}$.  For $\vp=0,\pi$, $\tg_{mn}$
corresponds to the metric of the four-manifold in the string
frame.  Here, $g_m$ is purely real and only the NS-NS 2-form
$B_{mn}$ is turned on (see (B.3)).  This is indeed the result for
NS 5-branes [\chsb].  As for $\vp=\pi/2\,, 3\pi/2$, $g_m$ is
purely imaginary and signifies the presence of $D5$- and/or ${\bar
D}$5-branes.  With sources, \dlap\ gets modified to
$$
\bbox D = \star \r_{5} \eqn\ddlap
$$
where $\r_5$ is the density distribution of the 5-branes.

The above special class of type IIB solution are in a sense two
copies of analogous six-dimensional heterotic backgrounds. It is
straightforward to show that the four-manifold must be conformally
Calabi-Yau in the heterotic case. In order to see this consider
the heterotic dilatino and gravitino supersymmetry constraints in
the string frame\foot{We use the standard notation in the
NS5-brane literature [\chs] which differs from that in [\andy] by
a rescaling of the dilaton.  Also, the string metric is related to
Einstein metric by $g_{MN}^E=e^{-\ph/ 2}\,g_{MN}^S$.}
$$
\d \l  = \G^M\pa_M\ph\, \e - {1\o 6} H_{PQR}\,\G^{PQR}\,\e=0 ~.
\eqn\hdila
$$
$$
\d \psi_M = \nabla_M \e - {1\o 4} H_{MPQ}\,\G^{PQ}\, \e=0 ~,
\eqn\hgra
$$
As shown in [\andy], the string frame metric must take the form
$$
ds^2=\eta_{\m\n}\, dx^\m dx^\n+ g_{mn}(y)\,dy^m dy^n~.
\eqn\hmetric
$$
We shall take $\ph=\ph(y)$ and let the only nonzero components of
the three-form take the form $H_{mnp}={\e_{mnp}}^qh_q(y)$. The
supersymmetry transformations become

$$
\d\l = \G^m\left(\pa_m\ph - h_m\, \G^{(4)}\right) \e =0 \eqn\hdilb
$$
$$
\d \psi_m = \left(\na_m - {1\o 2} {\G_m}^r\,h_r\,
\G^{(4)}\right)\e=0~. \eqn\hgrb
$$
We now rescale the internal metric, according to
$g_{mn}=e^{2\ph}g'_{mn}\,$ and find
$$
\na'_m\e =0~. \eqn\hgravb
$$
Therefore, $g'_{mn}$ can only be a Calabi-Yau metric and the
internal four-manifold is conformally Calabi-Yau.

\chapter{Conclusion and Outlook}

In this paper we have discussed the compactification of type IIB
string theory to six dimensions. With non-vanishing fluxes, we
have translated the conditions for unbroken supersymmetry into
conditions on the background geometry and the tensor fields. Our
work can be viewed as a step towards a complete classification of
string theory vacua in six dimensions.

There are several open questions which we will leave for future
work. First, we have focused on six-dimensional space-times with a
vanishing cosmological constant. In this case we have seen that
the conditions for unbroken supersymmetry factorize into two
independent conditions involving spinors of a definite
six-dimensional chirality only. An immediate open question is if
unbroken supersymmetry implies a vanishing of the six-dimensional
cosmological constant. If $AdS_6$ backgrounds exist they would
have the interesting property that the equations involving spinors
of different six-dimensional chirality do not decouple. Such a
property has been found in supergravity backgrounds like for
example in [\PolchinskiUF].

Another open question concerns compactifications of the heterotic
string to six dimensions. One of the most exciting string theory
developments in the last few years is the statistical approach to
string theory compactifications initiated by Douglas and
collaborators [\DouglasUM]. This approach opens the door to the
possibility of making real world predictions, maybe for the scale
of supersymmetry breaking or for the possibility that large extra
dimensions appear in nature. It would certainly be very
interesting to generalize this statistical approach to
compactifications of the heterotic string. Studying the
distribution and number of flux vacua of the heterotic string
compactified on a six-dimensional torsional background certainly
sounds like a formidable task since not much is known about vector
bundles on manifolds with torsion. But it would be an interesting
problem to study the number of heterotic flux vacua in six
dimensions. The background geometry can only be conformal to a
Calabi-Yau two-fold (either $K3$ or $T^4$). Moreover, the fluxes
are gauge fields and are constrained to satisfy the
Donaldson-Uhlenbeck-Yau equations
$$
F_{a b } = F_{\bar a \bar b} = F_{a \bar b} J^{a \bar b}=0~.
\eqn\duye
$$

In summary flux backgrounds in six dimensions are simple enough
that the background geometry can be described in a concrete way
yet complicated enough to capture many interesting properties. We
will return to the issues raised above in future works.

\vskip 1cm

\no{\bf Acknowledgements} \break \no We are grateful to Melanie
Becker, Keshav Dasgupta, Eric Sharpe, and Radu Tatar for
interesting discussions.  We also thank James Liu and Douglas
Smith for helpful conversations.  This work was supported in part
by NSF grant PHY-0244722, an Alfred P. Sloan fellowship and the
University of Utah.

\vfill

\break


\Appendix{A}

\no Our notation and conventions are as follows (a good reference
about spinors and properties of the Clifford algebra is
[\VanProeyenNI]):

\item{\tria} The different types of indices that we use are:
$$
\eqalign{ M, \, N, \dots \qquad &\hbox{are ten-dimensional
lorentzian indices},\cr A, \, B, \dots \qquad &\hbox{are
ten-dimensional tangent space indices},\cr \m, \; \nu, \dots
\qquad &\hbox{are six-dimensional lorentzian indices},\cr m,\;
n,\dots \qquad &\hbox{are real indices of the euclidean
submanifold},\cr a, \; b, \dots {\rm and} \; {\bar a}, \; {\bar
b}, \dots \qquad & \hbox{are complex indices of the euclidean
submanifold},\cr }
$$
In addition, we use $i,j,k,l=1,2$ as $SU(2)$ indices labelling
different spinors and not their components.  Moreover, the
coordinates of the external space are denoted by
$x=(x^0,x^1,\dots, x^5)$ while $y=(x^6, x^7, x^8, x^9)$ denotes
the coordinates of the four-manifold.

\item{\tria} We follow the mostly positive signature for our
metrics. The gamma-matrices ${\G}^A$ are hermitian, for $A=1,
\dots, 9$ while ${\G}^0$ is antihermitian. They satisfy
$$
\{ {\G}^A, {\G}^B \} =2{\et}^{AB}~, \eqn\algebra
$$
where ${\et}^{AB}$ has the signature $(-,+,\dots,+)$.  We
decompose the 10d gamma matrices as follows.
$$
\eqalign{ \G^A&=\g^A \ot \G^{(4)}~, \qquad A=0,\ldots,5 \cr
\G^A&=\I^{(6)}\ot\g^A~, \qquad A=6,\ldots,9 } \eqn\sfgamma
$$
where $\I^{(6)}$ is the 6d identity matrix and $\G^{(4)}$ is the
4d chirality matrix.  An explicit representation is given by
$$
\eqalign{ \g^0 &= i\s_2 \ot \s_3 \ot \s_3~, \cr \g^1 &=\s_1 \ot
\s_3 \ot \s_3~,}\qquad\eqalign{\g^2 &=\I \ot \s_1 \ot \s_3~,
\cr\g^3 &=\I \ot \s_2 \ot \s_3~,}\qquad\eqalign{\g^4 &= \I \ot \I
\ot \s_1~,\cr \g^5 &= \I \ot \I  \ot \s_2~,}
$$
$$
\eqalign{ \g^6 &= \s_1 \ot \s_3~, \cr\g^7 &= \s_2 \ot
\s_3~,}\qquad \eqalign{\g^8 &= \I \ot \s_1~,\cr \g^9 &= \I \ot
\s_2~,}
$$
where $\s_i$ are the Pauli matrices and $\I$ is the $2\times 2$
identity matrix.  The 10d chirality matrix is
$$
\eqalign{\G^{(10)} &= \G^0\G^1\cdots\G^9 \cr &= \s_3 \ot \s_3 \ot
\s_3 \ot \s_3 \ot \s_3 ~,} \eqn\chiral
$$
and can be written as $\G^{(10)} = \G^{(6)} \ot \G^{(4)}$ where
the 6d and 4d chirality matrices are defined as
$$
\eqalign{\G^{(6)}& = -\g^0\g^1\cdots\g^5 =\s_3 \ot \s_3 \ot \s_3
\cr \G^{(4)}&= -\g^6\g^7\g^8\g^9= \s_3 \ot \s_3~.} \eqn\sixfoch
$$

\item{\tria} The complex conjugation matrix, $B^{(10)}$, satisfies
$$
B^{(10)}\,\G^A (B^{(10)})^{-1} = - \G^{A\star}\quad {\rm
and~}\quad B^{(10)}\,\S^{AB}(B^{(10)})^{-1}=-\S^{AB\,\star}~,
\eqn\bcond
$$
where $\S^{AB}=-{i\o 4}[\G^A.\G^B]$ are Lorentz generators.   In
10d, $B^{(10)}$ is explicitly
$$
B^{(10)}=\G^{(10)}\G^3\G^5\G^7\G^9=\s_3\ot\s_1\ot i\s_2\ot\s_1\ot
i\s_2~. \eqn\bten
$$
Note that $B^{(10)\star} B^{(10)}=1$.  Furthermore, we can
decompose $B^{(10)}=B^{(6)}\ot B^{(4)}$ with
$$
\eqalign{ B^{(6)}&=\G^{(6)}\g^3\g^5=\s_3\ot\s_1\ot -i\s_2~,\cr
B^{(4)}&=\G^{(4)}\g^7\g^9=\s_1\ot -i\s_2~.} \eqn\bsf
$$
Notice that $B^{(6)\star} B^{(6)}=B^{(4)\star} B^{(4)}=-1$, and in
particular, ${B^{(4)\,T}}=-B^{(4)}$.

\item{\tria} With coordinate indices, $\e_{m_1\ldots m_d}$ denotes
the Levi-Civita tensor.  In particular, $\e_{6789}=\sqrt{|g|}$
with the metric referring to that in \metric.  However, with
indices labelling spinors, $\e_{ij}$ is defined with the values
$\e_{12}=\e^{12}=1$.

%
\item{\tria} We have also defined
$$
/\!\!\!\! H = { 1\o n!} H_{N_1 \dots N_n} \G^{N_1 \dots N_n}~.
\eqn\slashdef
$$

\item{\tria} Some useful gamma matrix identities are
$$
\g_{mn}\,\G^{(4)}= {1\o 2}\, \e_{mnpq}\, \g^{pq}~,\qquad \g_m
\G^{(4)}= - {1\o 6}\, \e_{mnpq} \g^{npq}~, \eqn\gai
$$
$$
\eqalign{
& [\g_{mn} ,\g^r ]=-4\,{\d^r}_{[m} \g_{n]}~, \cr & [\g_{mnp}, \g^r
]=2\,{ \g_{mnp}}^r~,\cr & [\g_{mn},\g^{pq}] = -4\,
{\d_{[m}}^{[p}\,{\g_{n]}}^{q]}~,
  }\qquad
\eqalign{
& \{\g_{mn}, \g^r \} = {2\,\g_{mn}}^r~,\cr & \{ \g_{mnp}, \g^r\}
=6\, {\d^r}_{[m} \g_{np]}~, \cr & \{ \g_{mn}, \g^{pq}\} = 2\,
{\g_{mn}}^{pq} - 4\,{\d_{[mn]}}^{pq}~.} \eqn\gaii
$$

\item{\tria}The 4d Fierz identity
$$
\chi \psi^\dg ={1\o 4}  \sum_{n=0}^4\, {1\o n!}\, \G^{c_n \dots
c_1}\, \psi^\dg \G_{c_1 \dots c_n} \chi~, \eqn\fierz
$$
can be used together with \gaii\ to derive the following useful
formula
$$
\eqalign{ {\l^{\dg\,i}}{\g_m}^n\l_j\,{\l^{\dg\, k}}{\g_n}^p\l_l &
={\l^{\dg\,i}}{\g_m}^p\l_l\,{\l^{\dg\,k}}\l_j -
{\l^{\dg\,i}}\l_l\,{\l^{\dg\,k}}{\g_m}^p\l_j \cr & \qquad +
{\d_m}^p\left(2\,{\l^{\dg\,i}}\l_l\,{\l^{\dg\,k}}\l_j
-{\l^{\dg\,i}}\l_j\,{\l^{\dg\,k}}\l_l\right)~.} \eqn\fform
$$

\Appendix{B} Some formulas and definitions of the fields in type
IIB supergravity [\schwarz].
$$\eqalign{
\t&=\t_1+i\t_2=C_0+ie^{-\ph}~,\cr B&=\frac{1+i\t}{1-i\t}~,\cr
f&=(1-B^*B)^{-\frac{1}{2}}~,\cr}\qquad\eqalign{ &P_M=f^2\pa_M
B~,\cr &Q_M=f^2 {\rm Im}\left(B\pa_M B^*\right)~,\cr
&G_{MNP}=f\left(F_{MNP}-BF^*_{MNP}\right)~.} \eqn\formuls
$$
$F_{MNP}$ is related to the NS-NS and R-R two-forms by
$F_3={g\o\k}(dB_2+ i\, dC_2)$ where $g^2=2\k^2/((2\pi)^7\a'^4)$
[\grapol].
If we take $\ta=C_0=0$, then the formulas reduce to
$$
\eqalign{ B&=\tanh\pot~,\cr
f&=\frac{1+\tb}{2\sqrt{\tb}}=\cosh\pot~,\cr}\qquad\qquad\eqalign{
P_M&=-\frac{\pa_M\tb}{2\tb}=\frac{1}{2}\pa_M\ph~,\cr
Q_M&=\frac{-1}{2\tb}{\rm
Im}\left[\pa_M\tb\left(\frac{1-\tb}{1+\tb}\right)\right]=0~,\cr}
\eqn\fromcz
$$
$$
\eqalign{G_{MNP}&={g\o\k}\left(f(1-B)\,dB_2+i\,f(1+B)\,dC_2\right)\cr&={g\o\k}\left(e^{-\pot}dB_2+i\,e^{\pot}dC_2\right)~.}\eqn\fromczz
$$

\vfill \break \refout
\end